\documentclass[aps, prl, reprint, amsmath, longbibliography,superscriptaddress]{revtex4-1}
\usepackage[T1]{fontenc}
\usepackage{hyperref}
\usepackage{graphicx}
\usepackage[capitalize]{cleveref}
\usepackage{xcolor}
\usepackage{comment}
\usepackage{verbatim} 
\usepackage{xr}

\UseRawInputEncoding
\begin{document}

\title{Entropic Trapping of Hard Spheres in Spherical Confinement}

\author{Praveen K. Bommineni} \email{praveen@nitw.ac.in}
\affiliation{Department of Chemical Engineering, National  Institute of Technology Warangal, 506004 Telangana, India}
\affiliation{Institute for Multiscale Simulation, Friedrich-Alexander-Universit\"at Erlangen-N\"urnberg, 91058 Erlangen, Germany}
\author{Junwei Wang} \email{jwwang@eitech.edu.cn}
\affiliation{Institute of Particle Technology, Friedrich-Alexander-Universit\"at Erlangen-N\"urnberg, 91058 Erlangen, Germany}
\author{Nicolas Vogel} \email{nicolas.vogel@fau.de}
\affiliation{Institute of Particle Technology, Friedrich-Alexander-Universit\"at Erlangen-N\"urnberg, 91058 Erlangen, Germany}
\author{Michael Engel} \email{michael.engel@fau.de}
\affiliation{Institute for Multiscale Simulation, Friedrich-Alexander-Universit\"at Erlangen-N\"urnberg,  91058 Erlangen, Germany}

\begin{abstract}
Monodisperse spherical colloidal particles confined within emulsion droplets can crystallize into icosahedral clusters.
Experimentally it was observed that a few large colloidal particles added as defects preferentially migrate to the vertices of the icoshedral clusters.
To understand this structure formation phenomenon, we simulate the confined self-assembly of hard spheres in the presence of a small number of larger particles.
The results demonstrate that large spheres are significantly influenced by concentric shells of small spheres near the crystallization transition.
Entropic forces drive the large spheres to the cluster surface, where they settle into free energy minima at the icosahedron vertices.
Notably, the addition of twelve large spheres results in the formation of a perfect icosahedral frame.
Free energy calculations via umbrella sampling are used to quantify this process and show that both the migration to the cluster surface and the trapping at the vertices with trapping strength of multiple $k_\text{B}T$ results from free energy minimization.
Moreover, our study reveals that the crystallization pathway and dynamics of large spheres are consistent across different systems, suggesting robustness of entropic trapping.
\end{abstract}
\maketitle

\textit{Introduction.}---Monodisperse colloidal particles in bulk systems reliably self-assemble into close-packed crystalline structures~\cite{Alder1957a,Auer2001}.
Yet, the presence of impurity particles with a different size can significantly influence the nucleation, growth, and final structure of these crystals. 
Depending on their size, a single spherical impurity can either facilitate crystallization~\cite{Cacciuto2004,Allahyarov2015} or introduce geometrical frustration that hinders it~\cite{Villeneuve2005}. 
Multiple impurity particles tend to segregate and accumulate at grain boundaries~\cite{Nozawa2013,Lavergne2016,Guo2018}. In the kinetic limit, when the drying process occurs at large Peclet numbers, diffusophoretic effects can cause an enrichment of smaller particles at the interface~\cite{Zellmer2015,Fortini2016,Liu2019}.
Confining elements, such as the spherical nature of a particle-ladden emulsion droplets impose geometric restrictions on the self-assembly process, which can cause the formation of different types of local and global order~\cite{Manoharan2003,DeNijs2015,Teich2016,Kister2016,Wang2018,Wang2018a,Montanarella2018,Wang2019,Kim2020,Wang2021a,Wang2021, Zhu2021,Lee2022,Mbah2023}.
A prominent example are confined monodisperse colloidal particles crystallizing into thermodynamically stable icosahedral clusters~\cite{DeNijs2015,Wang2018,Wang2019,Lee2022,Mbah2023}.
In these clusters, twenty tetrahedral grains meet at twelve line defects known as triple junctions, which intersect at the vertices of the icosahedron.
Such colloidal clusters also exhibit magic number phenomena, similar to those observed in atomic clusters~\cite{Wang2018}, evidenced by the tendency to form closed outer shells with high structural order~\cite{Lacava2012,Wang2018,Choi2024}.

In this study, we explore the behavior of colloidal clusters consisting of a few large hard spheres added to systems of small hard spheres. 
We show that the large spheres are driven to the crystal surface, similar to the diffusion of impurities to grain interfaces in nanoalloys~\cite{Nelli2021}, and subsequently migrate to the icosahedral vertices where they become trapped.
We calculate free energy profiles responsible for this entropic trapping effect using Monte Carlo umbrella sampling. Importantly, and in contrast to stratification effects occurring in binary mixtures drying at high Peclet numbers~\cite{Zellmer2015,Fortini2016,Liu2019}, the entropic trapping of particles at defined positions of the icosahedral cluster occurs during slow assembly, such that all particles can efficiently relax into minimum free energy positions.
Our work demonstrates the positioning of particles at specific locations and thus reveals fundamental structure formation pathways for the design of complex materials by self-assembly.

\begin{figure}
\includegraphics[width=0.95\linewidth]{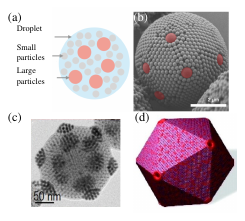}
\caption{Examples demonstrating templating of icosahedral clusters.
(a)~Schematic representation of a mixture of many small particles and a few large particles in an emulsion droplet.
(b)~Scanning electron microscopy observation of the successful diffusion of five large polystyrene colloidal particles (red) into the vertices of the icosahedral cluster.
(c)~Co-assembly of two types of grafted nanoparticles into a patchy icosahedron in a drying emulsion droplet~\cite{Meng2024}. Reprinted with permission from AAAS.
(d)~Molecular crystallization and segregation of anionic and cationic surfactants~\cite{Dubois2004}. Copyright 2004 by the National Academy of Sciences.
}
\label{fig:Fig1}
\end{figure}

\textit{Experimental motivation.}---This study begins with an experimental observation.
We mixed a large number of small charge-stabilized polystyrene colloidal particles (diameter 240~nm) with a relatively small number (ratio 500:1) of large polystyrene colloidal particles (diameter 1~{\textmu}m) and formed colloidal clusters by confined self-assembly within water-in-oil emulsion droplets~\cite{DeNijs2015,Wang2018} (\cref{fig:Fig1}(a)). 
At sufficiently slow drying speed, where the system can efficiently relax, well-equilibrated colloidal clusters with icosahedral symmetry are formed~\cite{DeNijs2015,Wang2018,Choi2024}.
When larger defect particles were present, they were reproducibly incorporated within the icosahedral vertices at the cluster surface (\cref{fig:Fig1}(b)).
It appears that  grain boundaries, particularly triple junctions, and most preferentially the ends of these junctions at the icosahedral vertices, trap the large particles.
Given that the interactions between our colloidal particles are essentially hard, we assume that this trapping effect is predominantly entropic in nature.
Similar icosahedral templating has recently been observed in binary mixtures of nanoparticles and surfactants (\cref{fig:Fig1}(c,d)), suggesting that this trapping phenomenon is more general and found in many types of systems with diverse interactions.

\begin{figure}
\includegraphics[width=\linewidth]{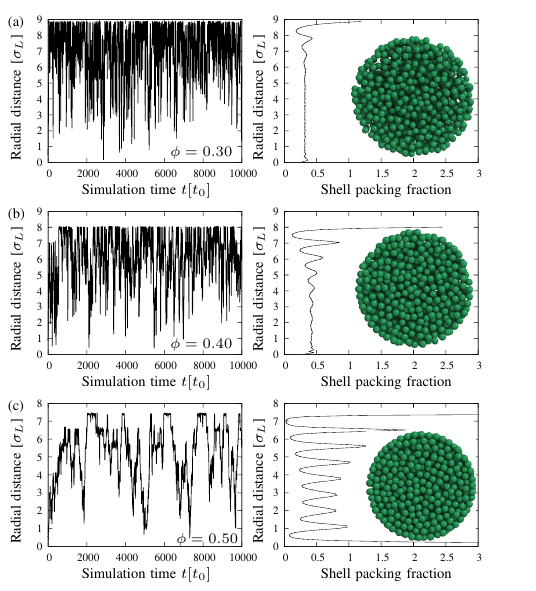}
\caption{Layering in confined hard sphere fluids.
The left side shows the evolution of radial distance from the confinement center of a randomly chosen test sphere, while the right side displays shell packing fraction profiles in thin spherical shells obtained by averaging over many frames.
Insets show typical cross-sectional views of the confined fluids.
Clusters are shown at three packing fractions: (a)~$\phi=0.30$, (b)~$\phi=0.40$, and (c)~$\phi=0.50$.
Note that layering persists throughout the entire cluster in (c), even though this cluster is not yet crystalline.
Simulations were performed with $N=2000$ spheres.}
\label{fig:Fig2}
\end{figure}

\textit{Layering in confinement.}---To understand the physics governing the experimentally observed migration effect, we systematically investigate the thermodynamics and dynamics of larger impurity particles in the self-assembly of hard spheres in spherical confinement using event-driven molecular dynamics (EDMD) simulations in the $NVT$ ensemble. 
The interactions between the particles, as well as between the spheres and the spherical confinement, are modeled using hard sphere potentials. 
The size ratio between the small and large spheres is defined as $\alpha=\sigma_S/\sigma_L$, where $\sigma_S$ and $\sigma_L$ are the diameters of the small and large spheres, respectively.

It is well-known that spheres near a confining interface behave differently than in bulk and can develop characteristic density fluctuations~\cite{Jianping1997, Mittal2008, Nygard2016}. This layering effect has been shown to influence the preferred packing in planar systems~\cite{Fortini2006,Mittal2008}, and to template icosahedral symmetry in curved confinements~\cite{DeNijs2015,Chen2021}.
To probe how these layering phenomena relate to the migration of larger defect particles, we first analyze density fluctuations in homogeneous systems by simulating identical hard spheres at three packing fractions.
We analyze density fluctuations with the help of a local measure, which quantifies the packing fraction in spherical shells and which we term shell packing fraction (see Supplemental Material).
At packing fraction $\phi=0.30$, the spheres diffuse nearly freely and shell packing fraction remains constant near the confinement center (\cref{fig:Fig2}(a)). 
As packing fraction increases to $\phi=0.40$, the mobility of the spheres decreases and the fluid forms layers near the cluster surface, as indicated by the presence of multiple shell packing fraction oscillations (\cref{fig:Fig2}(b)).
The highest packing fraction $\phi=0.50$ creates pronounced layering, with distinct steps in the sphere trajectories and shell packing fraction oscillations extending throughout the entire cluster (\cref{fig:Fig2}(c)). These layers subsequently template the crystallization into the final icosahedral symmetry~\cite{DeNijs2015,Chen2021}.

\begin{figure*}
\includegraphics[width=\linewidth]{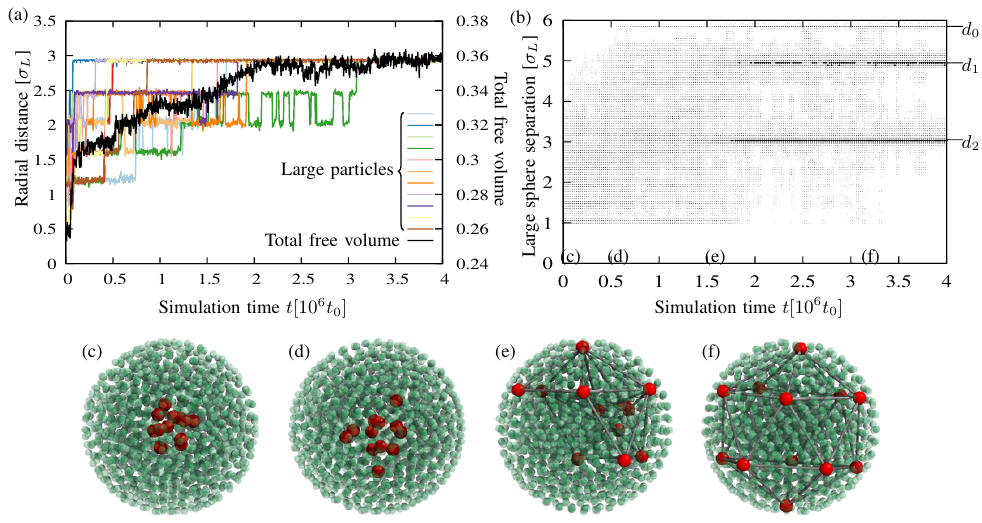}
\caption{Twelve large spheres form a perfect icosahedral frame.
(a)~Large spheres hop between shells increasing total free volume, which tracks crystallization progress.
(b)~Histogram showing the evolution of large sphere separations as they settle into the icosahedral frame.
Simulation snapshots show: (c)~Initial setup with large spheres positioned at the confinement center;
(d)~Outward diffusion of large spheres;
(e)~Intermediate state showing a partially formed icosahedral frame;
(f)~Completed icosahedral frame at the end of the trajectory.
EDMD simulations were conducted with $N_L=12$ large and $N_S=1288$ small spheres at fluid-solid coexisting packing fraction $\phi=0.53$ and for sphere size ratio $\alpha=0.55$.}
\label{fig:Fig3}
\end{figure*}

\textit{Migration of large particles to the icosahedral vertex positions.}---As we have seen, packing fraction strongly influences layering near the confining interface in a system of small spheres. We now
add a few large spheres to investigate their trajectory during the self-assembly process.
We choose a magic number configuration~\cite{Wang2018,Wang2019} for the majority population of small particles, $N_S=1288$, and add 12 larger particles, following the experimental observation that these preferentially occupy the 12 vertex position at the cluster surface. Packing fraction is slightly increased to $\phi=0.53$, where the layering behavior transitions into crystallization behavior.
We initialize a simulation with the small spheres in a fluid state and the large spheres positioned near the center of the confinement.
Two simultaneous ordering processes are observed: the small spheres crystallize into an icosahedral cluster, and the large spheres diffuse toward the cluster surface.
To demonstrate the latter, we track the radial distance from the confinement center of the twelve large spheres (\cref{fig:Fig3}(a)).
The large spheres quickly move out of the confinement center and gradually hop from shell to shell.
The large spheres tend to remain in the center of the shells rather than between layers, which demonstrates that outwards diffusion is strongly affected by the layering.
Once the large spheres reach the outermost shell, they remain confined to it. 

To understand the driving force behind the outward diffusion of the large spheres, we calculated the total free volume along the simulation trajectory.
The total free volume refers to the space available for any sphere to move while all other spheres remain fixed; its evolution approximates the free energy change of the system~\cite{Hoover1968,Haji-Akbari2011}.
As shown in \cref{fig:Fig3}(a), the initially disordered cluster has low total free volume.
Free volume gradually increases while the small spheres crystallize and the large spheres diffuse outward.
As shown in Fig.~S1~\footnote{See Supplemental Material for Figures S1 to S5, a detailed description of methods, which includes Refs.~\cite{Bommineni2019, Bommineni2020, Isobe_2016}.}, pressure decreases along the simulation trajectory, and mean squared displacement captures the diffusion and the eventual arrest of the large spheres.

Simultaneously to the large spheres migrating to the confinement surface, they become influenced by grain boundaries, which guide them towards specific sites on the confinement surface at the end of triple junctions.
We track the arrangement of the large spheres by calculating the histogram of their spatial separation in \cref{fig:Fig3}(b).
The size of the dots in the histogram corresponds to the probability of a certain separation to be found in the system.
Peaks in the histogram develop at separations $d_0=5.9$, $d_1=5.0$, and 
$d_2=3.1$, where $d_0$ is the distance between antipodal points.
The ratios of these distances match the ideal values for an icosahedron, $d_1/d_0=(\tau / \sqrt{5})^{1/2}$ and $d_1/d_2=\tau$, where $\tau=(\sqrt{5}+1)/2$ is the golden mean, confirming that the large spheres strive to form an icosahedral frame.
The formation of the icosahedral frame is also captured by simulation snapshots in \cref{fig:Fig3}(c-f) and in Movie~S1.

\begin{figure*}
\includegraphics[width=\linewidth]{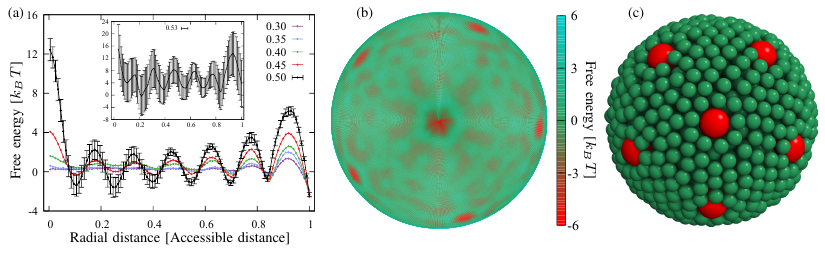}
\caption{Quantification of entropic trapping.
(a)~Free energy profiles of a single large test particle in a system of $N_S=1299$ small spheres as a function of radial distance.
Radial distance is measured from the confinement center and normalized by the distance accessible to the test sphere.
Simulation parameters are $\alpha=0.55$ and $\phi\in [0.30,0.35,\ldots,0.50]$ (labelled on the right); the inset corresponds to $\phi=0.53$.
For $\phi\geq 0.5$, we performed ten simulations for each biasing bin to improve statistics.
Solid lines are a guide to the eye.
Error bars indicate standard errors.
(b)~Two-dimensional free energy profile on the cluster surface.
(c)~Simulated cluster with twelve large spheres (red) oriented similar to the free energy profile. 
Large spheres are on the surface at the icosahedral vertices of the crystalline cluster formed by small spheres. 
}
\label{fig:Fig4}
\end{figure*}

\textit{Quantifying entropic trapping.}---Our simulations revealed the presence of an entropic force that drives the large spheres first to the confinement surface through the layer of shells formed by small spheres and then via surface diffusion to specific trap positions. 
To quantify this entropic driving force, we perform Monte Carlo simulations in the $NVT$ ensemble of clusters with one large sphere serving as a test sphere and calculate free energy using umbrella sampling with appropriately chosen reaction coordinates to analyze the two steps of the icosahedral frame formation: (i)~radial diffusion of large spheres to the confinement surface and (ii)~surface diffusion to trap positions.

In the first step, we analyze radial diffusion.
We perform umbrella sampling with an external harmonic bias potential of the form $U_i(\lambda)={\frac{1}{2}} k_\lambda (\lambda - \lambda_{i})^2$, where $k_\lambda=600k_\text{B}T/\sigma_L^2$ is a suitably chosen spring constant, $\lambda$ is the radial distance of the test sphere from the confinement center and $\lambda_{i}$ is the center of the $i$-th biasing window (Figs.~S2-S3).
We simulate each window for $10^5$ Monte Carlo cycles, where a cycle is defined as $N$ sphere moves.
Estimated biased probabilities $P_i(\lambda)$ from histogram data are unbiased using the weighted histogram analysis method (WHAM)~\cite{Ferrenberg1989,Kumar1992,Grossfield}.
Free energy profiles are obtained by normalizing unbiased probabilities with a volume correction term (Jacobian determinant~\cite{Johannes2011}) for each bin as $F_i(\lambda) / k_\text{B}T  = -\log (P_i(\lambda)V/V_i)$, where $V$ is the total confinement volume and $V_i$ is the volume of the $i$-th bin. 
The results are shown in \cref{fig:Fig4}(a) for packing fractions ranging from low ($\phi=0.30$) to high ($\phi=0.53$).
At all packing fractions, free energy profiles have a global minimum at the confinement surface, quantifying the outward diffusion of the large sphere.
Oscillations in the free energy profiles indicate the gradual appearance of layering in the small sphere fluid.
Layering hinders the diffusion of the test sphere until it becomes trapped at the surface.
At $\phi\geq 0.45$, layering first reaches the confinement center.
At $\phi=0.50$, just before crystallization, yet still in the fluid regime, layering is now fully developed with fluctuations in the free energy profile exceeding $4k_\text{B}T$.
At the crystallization packing fraction $\phi=0.53$ (inset of \cref{fig:Fig4}(a)), free energy fluctuations are more pronounced and difficult to sample, but indicate strong layering.

In the second step, we analyze angular diffusion to quantify the strength of the entropic traps with icosahedral geometry on the surface.
We estimate biased probabilities on the cluster surface using the biasing potential with spherical coordinates, $U_i(\theta,\varphi)=\frac{1}{2} k_\theta(\theta-\theta_i)^2 + \frac{1}{2}k_\varphi(\varphi-\varphi_i)^2$, where $k_\theta=k_\varphi=\frac{1}{2}k_\text{B}T$.
The reaction coordinates $\theta$ and $\varphi$ determine the position of the test sphere on the cluster surface, and $\theta_i$ and $\varphi_i$ are the centers of the $i$-th biasing window.
Simulations are performed with an equilibrated cluster configuration at $\phi=0.53$.
We normalize unbiased probabilities $P_i(\theta,\varphi)$ with a surface area correction term for each bin as $F_i(\theta,\varphi) / k_\text{B}T =-\log(P_i(\theta,\varphi) A/A_i)$, where $A$ is the total surface area of the cluster and $A_i$ is the area of the $i$-th bin.
The free energy profile on the cluster surface (\cref{fig:Fig4}(b)) exhibits deep minima (red patches) at the locations of the icosahedron vertices (\cref{fig:Fig4}(c)) with a trapping strength of about $|\Delta F_\text{trap}| = 6k_\text{B}T$, confirming the entropic trapping effect found in unbiased simulations.
Other shallower free energy minima on the surface indicate an intricate interplay between the geometry of the icosahedral cluster and the large test sphere.

\textit{Discussion.}---The icosahedral vertices of colloidal clusters provide space to accommodate impurity particles, such as the large spheres in our simulations. 
We assess the robustness of entropic trapping by simulating two more systems with the parameter pairs $(N,\alpha)=(2000,0.50)$ and $(2700,0.40)$.
The number of small spheres in these parameters are near magic numbers.
We track the radial distances of twelve large spheres from the confinement center and examine histograms of large sphere separations.
The trajectories closely mirror the patterns in~\cref{fig:Fig3}(a,b): 
Large spheres initially hop between shells formed by small spheres, progressing toward the confinement surface through radial diffusion and ultimately settled into the icosahedral vertices~(Figs.~S4-S5).
We repeat simulations three times for both parameter sets and observe entropic trapping in all simulations.
Such consistency across different systems demonstrates that entropic trapping is a robust phenomenon.

Our findings contribute to the growing body of research that underscores the pivotal role of geometry in particle assembly and trapping within confined systems~\cite{Dinsmore2002, Francisco2016, Chen2021, Das2022, Hacmon2023, Meng2024, Wan2024, zhu2024, Canestrari2025}.
By carefully controlling variables, such as the number of large spheres and the size and shape of the confinement, various patterning effects~\cite{Li2021} can be achieved.
The ability to direct objects to specific places in a confined system shows how structural complexity can be encoded within a self-assembling system. Our findings establish the physics governing the experimentally observed particle trapping in icosahedral clusters and can inform the design of advanced materials. Parallels between entropically trapping in particle assembly and the complex structure of icosahedral virus capsids and protein complex~\cite{Zlotnick_1994, Hagan_2014, Bale_2016} may suggest that similar physics may be at play in structure formation processes in biological systems as well.

\begin{acknowledgments}
\textit{Acknowledgments.}---We acknowledge support by the Deutsche Forschungsgemeinschaft (DFG, German Research Foundation) under Project-IDs 338276051 (EN 905/2-1 and VO 1824/7-1) and 416229255 (SFB 1411).
PB acknowledges NIT Warangal and SFB 1411 for enabling a sabbatical stay at FAU Erlangen.
HPC resources provided by the Erlangen National High Performance Computing Center (NHR@FAU) under the NHR project CRC1411D04 are gratefully acknowledged.
\end{acknowledgments}

%


\clearpage
\setcounter{figure}{0}
\renewcommand{\thefigure}{S\arabic{figure}}
\onecolumngrid

\section{SUPPLEMENTAL MATERIAL}
\subsection{Event-Driven Molecular Dynamics Simulations}

Simulations that revealed entropic trapping phenomena in binary mixtures of large and small hard spheres in hard spherical confinement are performed using the event-driven molecular dynamics (EDMD) method in the $NVT$ ensemble~\cite{Wang2018,Bommineni2019,Bommineni2020}. 
The total number of particles is $N=N_L+N_S$, where $N_L$ is number of large spheres and $N_S$ is number of small spheres. 
Size ratio of spheres is defined as $\alpha = \sigma_S / \sigma_L$, where $\sigma_S$ and $\sigma_L$ are diameters of small and large spheres. Masses of the spheres are set equal, $m_L = m_S = m$. 

The packing fraction of the system is given by $ \phi =  (N_L\sigma_L^3 + N_S \sigma_S^3)/D^3$, where $D$ is diameter of the confinement sphere.
Simulations are run for $4\times10^6t_0$ with the unit of time $t_0=\sigma_L \sqrt{m/{k_BT}}$.
$k_BT$ is the thermal energy. 
Along the simulation trajectories, we measured mean squared displacement (MSD) of large spheres and dimensionless pressure.
The pressure calculation is detailed in the last section of this document.
We analyze the radial dependence of local density with the help of a quantity termed shell packing fraction using discrete bins.
The $i$-th bin comprises the spherical shell with radius range $r\in[ R_i,R_{i+1}]$.
Shell packing fraction in the $i$-th bin is defined as $\rho_i = \frac{n_i V_0}{V_i}$, where $n_i$ is the time-averaged number of particles in the bin, $V_0=\frac{4}{3} \pi (\sigma/2)^3$ the particle volume, and $V_i =  \frac{4}{3} \pi ({R_{i+1}^3 - R_{i}^3})$ the bin volume.
Note that shell packing fraction can be above 1, which is not possible for packing fraction.

\begin{figure}
\includegraphics[width=0.4\linewidth]{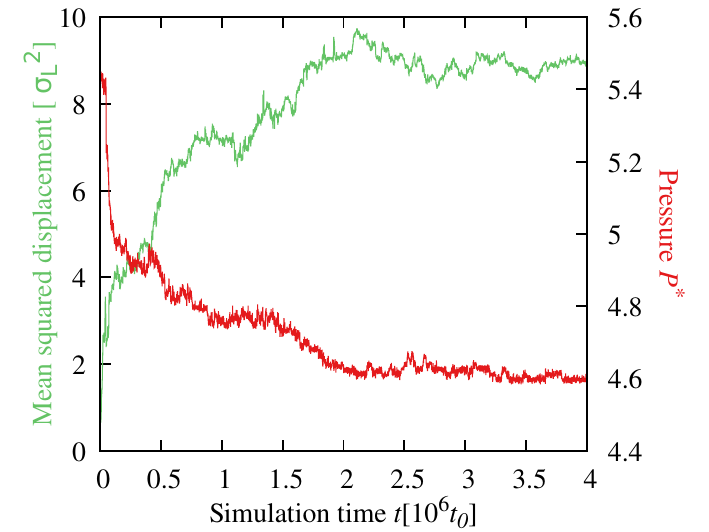}
\caption{Evolution of dimensionless pressure $P^{*}$ and mean squared displacement of large spheres with simulation time $t$ at the parameter values $N=1300, \alpha =0.55$ and $\phi=0.53$ identical to those in Fig.~3. $P^{*}$  gradually decreases during the crystallization and eventually stabilizes.}
\label{fig:SM0}
\end{figure}

\cref{fig:SM0} shows the evolution of dimensionless pressure and MSD with simulation time at the parameter set $(N,\alpha)=(1300,0.55)$ as in Fig.~3.
As the small spheres crystallize into an icosahedral cluster, $P^{*}$ decreases and remains constant after $t=2\times10^6t_0$.
Constant $P^{*}$ indicates completion of crystallization process. 
During this process, MSD of large spheres gradually increases as they diffuse to the surface and stabilizes as they settle into icosahedral cluster vertices formed by small spheres. 
The increase in MSD coincides with the increase of free volume measured in Fig.~3(a). 

\subsection{Free Volume Calculations}

Free volume of a particle is the measure of space available for a particle to move without overlapping with neighboring particles. 
Total free volume of system of $N$ particles is obtained by averaging free volumes of all particles. 
To measure the free volume of a particle, we use the binning method~\cite{Haji-Akbari2011}. 
In this method, $n_\text{bins}$ are constructed around a particle radially with volume $v_\text{bin}$. 
Then $n_\text{trial}$ moves are performed per bin and number of non-overlapping moves $n_\text{no}$ are recorded. 
The free volume is then calculated by using 
\begin{equation}
    v_\text{f} = \frac{1}{n_\text{trial}}\sum_{i=1}^{n_\text{bins}}v_{i,\text{bin}} n_{i,\text{no}}.
\end{equation}

\begin{figure}
\includegraphics[width=0.4\linewidth]{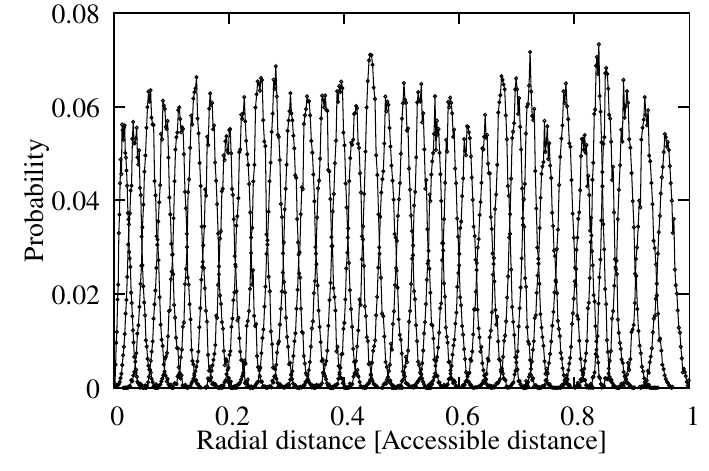}
\caption{Overlapping histograms obtained from umbrella sampling simulations for radial free energy profiles at $\phi=0.40$ with biasing constant $k=600k_BT$.}
\label{fig:SM4}
\end{figure}

\begin{figure}
\includegraphics[width=0.4\linewidth]{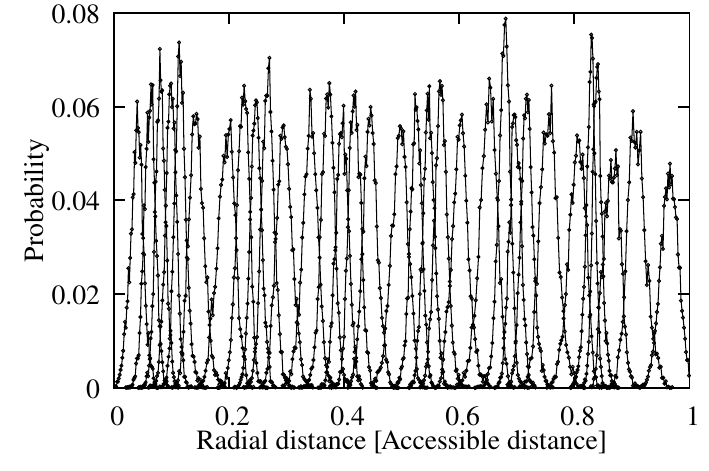}
\caption{Overlapping histograms obtained from umbrella sampling simulations for radial free energy profiles at $\phi=0.50$ using biasing constant $k=600k_BT$.}
\label{fig:SM5}
\end{figure}

\subsection{Free Energy Calculation via Umbrella Sampling}

To quantify trapping strength of icosahedral vertices, free energy calculations are performed in $NVT$ Monte Carlo simulations.
Radial free energy profiles and free energy landscape on the sphere surface are computed using the umbrella sampling method. 
For radial free energy profiles, a harmonic biasing potential of the form
\begin{equation}
    U_i(\lambda)= \frac{1}{2}k(\lambda-\lambda_i)^2
\end{equation}
is used, where $k$ is biasing constant, $\lambda$ is reaction co-ordinate that measures the distance of large sphere from the confinement center, and $\lambda_i$ is the center of $i$-th biasing bin. 
A biasing constant of $k=600k_BT$ is used for all biasing bins and $\lambda_i=0.1,0.2,0.3,\ldots$ are chosen for sampling in adjacent bins. 
This choice of parameters enables sufficient overlapping histograms as shown in \cref{fig:SM4,fig:SM5} for $\phi=0.40$ and $\phi=0.50$.
The Free energy profiles obtained by unbiasing estimated biased probabilities $P_i(\lambda)$ using weighted histogram analysis method (WHAM)~\cite{Ferrenberg1989,Kumar1992,Grossfield}.
Free energies are corrected by normalizing the unbiased probabilities with a Jacobian determinant~\cite{Johannes2011}.
The final corrected radial free energy is $F_i(\lambda) / k_\text{B}T  = -\log (P_i(\lambda)V/V_i) $ where the accessible volume for test sphere is $V= \frac{4}{3} \pi R^3$ and the volume of i-th bin is $V_i= \frac{4}{3} \pi (R_{i+1}^3-R_i^3)$.

For particles moving on the confinement sphere surface, free energy calculations of the test sphere are performed in spherical coordinates $(\theta,\varphi)$, which allows angular diffusion.
The two-dimensional biasing potential used is
\begin{equation}
    U_i(\theta,\varphi)= \frac{1}{2} k_{\theta} (\theta-\theta_i)^2 + \frac{1}{2}k_{\varphi}(\varphi-\varphi_i)^2,
\end{equation}
where $k_{\theta} = k_{\varphi} = \frac{1}{2}k_BT$.
$\theta$  and $\varphi$ are reaction co-ordinates that determine position of test sphere on the cluster surface.
The surface free energy landscape is obtained by unbiasing probabilities $P_i(\theta,\varphi)$ using WHAM.
Free energies are corrected by normalising unbiased probabilities.
The corrected surface free energy is $F_i(\lambda) / k_\text{B}T  = -\log (P_i(\lambda)A/A_i) $ where $A = 4\pi R^2$ and the area of surface patch is estimated as $A_i = R^2 (\theta_{i+1}-\theta_i) (\cos(\varphi_{i+1})-\cos(\varphi_i))$.

\subsection{Simulation Trajectories}
\begin{figure*}
\includegraphics[width=\linewidth]{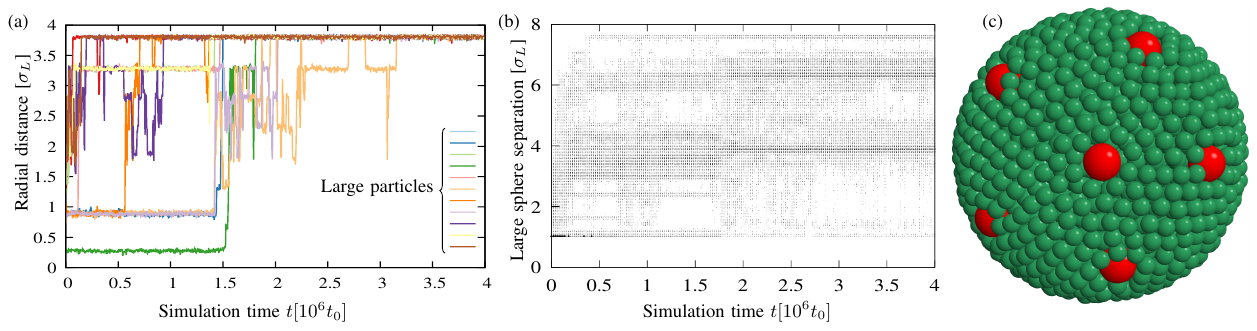}
\caption{ Evolution of twelve large spheres $N_L=12$ in a system of $N=2000$ at packing fraction $\phi =0.53$ and size ratio $\alpha = 0.50$. (a) Radial distances of large spheres measured from center of confinement, showing that spheres hop between shells. (b) Histogram of large sphere separations, initially the distances are small and increases over time as large spheres settle into icosahedral vertices . (c) Snapshot showing large spheres in the vertices of icosahedral cluster.}
\label{fig:SM2}
\end{figure*}

\begin{figure*}
\includegraphics[width=\linewidth]{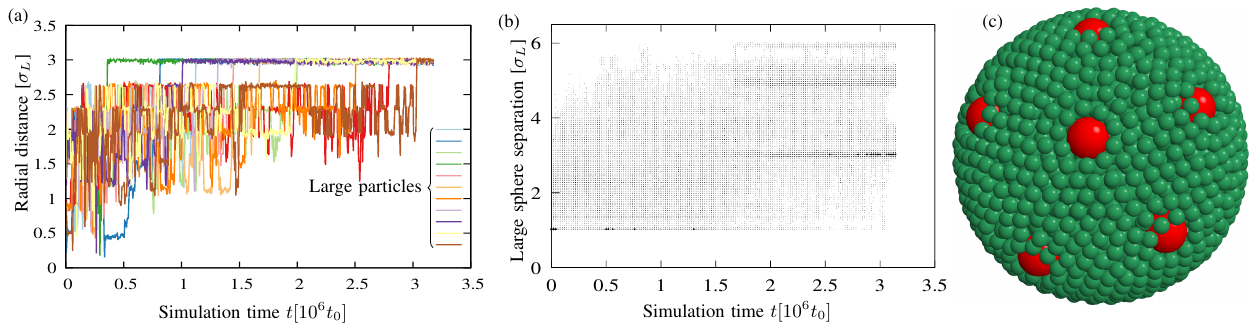}
\caption{Evolution of twelve large spheres $N_L=12$ in a system of $N=2700$ at packing fraction $\phi =0.53$ and size ratio $\alpha = 0.40$. (a) Radial distances of large spheres measured from center of confinement, showing that spheres hop between shells. (b) Histogram of large sphere separations, initially the distances are small and increases over time as large spheres settle into icosahedral vertices . (c) Snapshot showing large spheres in the vertices of icosahedral cluster.}
\label{fig:SM3}
\end{figure*}

Simulation trajectories discussed in main text with the parameter pairs $(N,\alpha)=(2000,0.50)$ and $(2700,0.40)$ are shown in \cref{fig:SM2,fig:SM3} respectively.


\subsection{Pressure calculation}

While the pressure formula for hard spheres exists in the literature~\cite{Isobe_2016}, derivations for a pressure formula for hard spheres interacting with a hard wall is not well covered. We derive the pressure formula for hard sphere mixtures assuming elastic collisions.

The derivation starts from the well-known virial pressure equation for soft particles,
\begin{equation}
P= \frac{N k_B T}{V} + \frac{1}{3V} \left\langle \sum_{i<j} \mathbf{r}_{ij} \cdot \mathbf{F}_{ij} \right\rangle.
\end{equation}
In the case of a binary mixture (generalization to other mixtures is straightforward), we define dimensionless pressure as
\begin{equation}
P^*=\frac{P \pi(\frac{N_L}{N}\sigma_L^3 + \frac{N_S}{N} \sigma_S^3)}{6k_B T}= \phi\left(1 + \frac{1}{3N k_B T} \left\langle \sum_{i<j} \mathbf{r}_{ij} \cdot \mathbf{F}_{ij} \right\rangle\right).
\end{equation}
To apply this equation to hard particles, we approximate all pair potentials with steep, repulsive potentials.
Such 'nearly-hard' particles undergo collisions with brief collision times.
In the limit of infinitely steep potentials, and thus infinitely brief collision times, all interactions decouple and collisions occur independently.
This allows us to replace the ensemble average by an average over collisions,
\begin{equation}
P^*=\phi\left(1 + \frac{1}{3N k_B T} \frac{1}{t_\text{tot}} \sum_\text{collisions} b_{ij} \right),
\end{equation}
where $t_\text{tot}$ is the total simulation time during which the collisions occur.
We call
\begin{equation}
b_{ij}=\int_{t'-\Delta t/2}^{t'+\Delta t/2} \mathbf{r}_{ij}(t) \cdot \mathbf{F}_{ij}(t)\;dt
\end{equation}
the collision action for the collision between particle $i$ and particle $j$.
It measures the total contribution of the collision to the virial pressure during the brief collision time $\Delta t$ in which the particles interact.
If there exist multiple distinct types of collisions, we can further separate the contributions and write
\begin{equation}
P^*=\phi\left(1 + \frac{1}{3N k_B T} \frac{1}{t_\text{tot}} \sum_C N_C \langle b_{ij} \rangle_C \right),
\end{equation}
where $\langle b_{ij} \rangle_C$ is the average collision action for collisions of type $C$.
Estimating pressure requires calculating the average collision action for each type of collision.

\subsection{Particle-wall collision}
First, we consider the collision of a sphere with diameter $\sigma$ and mass $m$ with a wall.
To keep the notation concise, particle index $j$ considers the wall as a hard `particle' that does not move.
For spherical particles, only the normal force $F^\perp_{ij}$ matters and the collision occurs at a fixed collision distance $\sigma/2$.
This simplifies the collision action to
\begin{equation}
b_{ij}=\frac{\sigma}{2}\int F^\perp_{ij}(t)\;dt=\sigma m v
\end{equation}
Here, we utilized that the integrated normal force is equal to a momentum transfer, where $v$ is the normal velocity right before the collision.
While particle velocities are distributed according to the Maxwell-Boltzmann probability distribution in equilibrium, faster particles collide more frequently with the wall.
Only particles moving with positive velocity (towards the wall) collide.
This means the probability distribution of the normal velocity right before the collision is given by
\begin{equation}
f(v)\propto \begin{cases}
v \exp\left(-\frac{mv^2}{2k_B T}\right) & v \geq 0, \\
0 & \text{else}.
\end{cases}
\end{equation}
The average collision action for a wall collision is given by
\begin{equation}
\langle b_{ij} \rangle=\sigma m \frac{\int_{-\infty}^\infty v f(v)\;dv}{\int_{-\infty}^\infty f(v)\;dv}=\sigma \sqrt{m k_B T} \sqrt{\frac{\pi}{2}}.
\end{equation}
Note that the shape of the wall was not relevant in the derivation.

\subsection{Symmetric particle-particle collision}
Next, we consider the collision of two sphere of identical diameter $\sigma$ and identical mass $m$.
Consider particle 1 and 2 have normal velocities $v_1$ and $v_2$ right before the collision. 
The collision occurs at a collision distance $\sigma$.
Because only the momentum transfer is relevant, we can use the center of mass frame.
The collision action is
\begin{equation}
b_{ij}=\sigma\int F^\perp_{ij}(t)\;dt=\sigma m (v_1 - v_2).
\end{equation}
Particles with faster relative velocity collide more frequently.
Only particles moving with positive relative velocity (towards each other) collide.
The probability distribution of the normal velocity right before the collision is given by
\begin{equation}
f(v_1,v_2)\propto \begin{cases}
(v_1 - v_2) \exp\left(-\frac{m v_1^2}{2 k_B T}\right)\exp\left(-\frac{m v_2 ^2}{2k_B T}\right) & v_1 > v_2, \\
0 & \text{else}.
\end{cases}
\end{equation}
The average collision action for the particle-particle collision is given by
\begin{equation}
\langle b_{ij} \rangle=\sigma m \frac{\int_{-\infty}^\infty\int_{-\infty}^\infty (v_1-v_2) f(v_1,v_2)\;dv_1dv_2} {\int_{-\infty}^\infty\int_{-\infty}^\infty f(v_1,v_2)\;dv_1dv_2} = \sigma \sqrt{m k_B T} \sqrt{\pi}.
\end{equation}
In the last step, we decoupled the double integral by introducing the new coordinates $v_1'=(v_1+v_2)/\sqrt{2}$ and $v_2'=(v_1-v_2)/\sqrt{2}$. This introduced a factor $\sqrt{2}$, which was not present in the particle-wall collision case.

\subsection{Asymmetric particle-particle collision}
Finally, we consider the most general case, the collision of two sphere of different diameters $\sigma_1$ and $\sigma_2$ and different masses $m_1$ and $m_2$.
Consider particle 1 and 2 have normal velocities $v_1$ and $v_2$ right before the collision. 
The collision occurs at a collision distance $(\sigma_1+\sigma_2)/2$.
Because only the momentum transfer is relevant, we can use the center of mass frame.
The collision action is
\begin{equation}
b_{ij}=\sigma\int F^\perp_{ij}(t)\;dt=\frac{\sigma_1+\sigma_2}{2} \frac{2m_1m_2}{m_1+m_2} (v_1 - v_2).
\end{equation}
Particles with faster relative velocity collide more frequently.
Only particles moving with positive relative velocity (towards each other) collide.
The probability distribution of the normal velocity right before the collision is given by
\begin{equation}
f(v_1,v_2)\propto \begin{cases}
(v_1 - v_2) \exp\left(-\frac{m_1 v_1^2}{2 k_B T}\right)\exp\left(-\frac{m_2 v_2 ^2}{2k_B T}\right) & v_1 > v_2, \\
0 & \text{else}.
\end{cases}
\end{equation}
The average collision action for the particle-particle collision is given by
\begin{equation}
\langle b_{ij} \rangle=(\sigma_1+\sigma_2) \frac{m_1m_2}{m_1+m_2} \frac{\int_{-\infty}^\infty\int_{-\infty}^\infty (v_1-v_2) f(v_1,v_2)\;dv_1dv_2} {\int_{-\infty}^\infty\int_{-\infty}^\infty f(v_1,v_2)\;dv_1dv_2}.
\end{equation}
This double integral is difficult to solve for general choices of the masses.
However, in the symmetric case $m=m_1=m_2$ considered in this manuscript, we can again introduce two new coordinates $v_1'=(v_1+v_2)/\sqrt{2}$ and $v_2'=(v_1-v_2)/\sqrt{2}$ and obtain
\begin{equation}
\langle b_{ij} \rangle=\frac{\sigma_1+\sigma_2}{2}\sqrt{m k_B T} \sqrt{\pi}.
\end{equation}
For $\sigma=\sigma_1=\sigma_2$ the symmetric particle-particle collision case is confirmed.

\subsection{Application to our binary system}
Our binary system contains three types of particles, small sphere ($S$), large sphere ($L$) and wall ($W$). There are five types of collision, $C\in[SS, LS, LL, SW, LW]$.
Combining all the derivations above, and setting $k_B T = m = 1$, the general virial pressure equation in dimensionless units reads
\begin{equation}
P^*=\phi\left(1 + \frac{\sqrt{\pi}}{3} \frac{1}{Nt^*_\text{tot}}
\left(\frac{N_{SS} \sigma_S + N_{LS} (\sigma_S + \sigma_L) / 2+ N_{LL} \sigma_L}{\sigma_L} +
\frac{N_{SW} \sigma_S + N_{LW} \sigma_L}{\sqrt{2}\sigma_L}
\right)\right).
\end{equation}
Here, $N_C$ are the number of collisions of type $C$ occurring during the dimensionless total simulation time $t^*_\text{tot}$.

\end{document}